\documentclass[aps,prl,10pt,twocolumn]{revtex4-1}
\usepackage{graphicx}
\usepackage{amsmath}
\usepackage{amssymb}
\usepackage{color}
\usepackage{mathrsfs}

\begin{document}


\title{Quantum Path Control in High-Order Harmonic Generation via Squeezed Lights}

\author{Feng Wang$^{1,*}$, Chunhui Yang$^{1,*}$, Xinyi Cui$^{2}$}\email{These authors contributed equally to this work.}

 \author{Lixin He$^{2}$}\email{helx\_hust@hust.edu.cn}

 \author{Tianxin Ou$^{1}$, Rui-Bo Jin$^{1}$, Qing Liao$^{1}$}
 \author{Pengfei Lan$^{2}$}\email{pengfeilan@hust.edu.cn}
 \author{Peixiang Lu$^{1,2,}$}\email{lupeixiang@hust.edu.cn}

\affiliation{%
 $^1$Hubei Key Laboratory of Optical Information and Pattern Recognition, Wuhan Institute of Technology, Wuhan 430205, China\\
 $^2$Wuhan National Laboratory for Optoelectronics and School of Physics, Huazhong University of Science and Technology, Wuhan 430074, China\\
}%

\date{\today}

\begin{abstract}
High-order harmonic generation (HHG), a robust tabletop source for producing attosecond pulses, has been extensively utilized in attosecond metrology. Traditionally, HHG driven by classical laser fields involves two typical quantum paths (short and long quantum paths) contributing to harmonic emission. Here, we demonstrate that these quantum paths in HHG can be selectively controlled using squeezed lights, a form of non-classical light. Our results indicate that the long (short) quantum path of HHG will be dramatically suppressed  in the  phase (amplitude)-squeezed fields. The time-frequency analysis reveals that this quantum path control stems from the quantum fluctuations in the squeezed light, which modify the phase matching of harmonic emission from different quantum states of the squeezed light. Such a quantum path selection can be achieved for the whole harmonic plateau, which has great potential to generate ultrashort isolated attosecond pulse with duration less than one atomic unit of time.  
\end{abstract}


\maketitle

High-order harmonic generation (HHG) is a highly nonlinear and nonperturbative process that occurs when intense laser fields interact with gases \cite{Krause}, solids \cite{Ghimire} or liquids \cite{Luu}.
Due to its coherent nature and broad plateau structure, HHG has emerged as a pivotal tabletop source for attosecond pulse generation \cite{Villeneuve,Hentschel,zhao1,Shirozhan}.
Attosecond pulses, with their unmatched temporal and spatial resolution, are essential in attosecond metrology \cite{Schultze,Dudovich,Grundmann,klunder,Paul} and ultrafast electronic dynamic imaging \cite{Smirnova,Worner,Lixin,AP,Liang1}.
In gases, the HHG process is well described by the semiclassical three-step model \cite{Corkum,Kulander}, which divides laser-gas interaction into ionization, acceleration and recombination steps.
Building on this, a quantum theory based on the strong-field approximation (SFA) \cite{Lewenstein} has been developed, where the harmonic dipole moment is expressed as a coherent sum of different quantum paths contributing to the HHG process.
For each harmonic below the cutoff, two main quantum paths, which are commonly referred to as the short and long paths, dominate the harmonic emission.
Control over these quantum paths is crucial in strong-field physics and attosecond science, enabling isolated attosecond pulse (IAP) generation \cite{Krausz,Worner1,53as,Chang1,80as,PG1,PR1,PR2,PR3,PR4}, time-resolved studies of molecular dynamics using high harmonic spectroscopy (HHS) \cite{Lein,Lein1,Peng,He2024}.


The HHG process is generally driven by intense laser pulse with large photon numbers, which is typically treated as a classical field with the quantum nature of the driving light completely neglected.
Recent advancements in laser technology, however, have enabled the experimental generation of intense quantum light sources.
For instance, bright squeezed vacuum (BSV) pulses with picosecond \cite{Iskhakov} and femtosecond \cite{Finger} duration have been experimentally demonstrated with energies up to 10 $\mu$J and 350 nJ, respectively.
The intensities of these quantum light sources are now approaching the strong-field regime, spurring significant interest in the intersection of strong-field physics and quantum optics.
Quantum lights offer new degrees of freedom, giving rise to some unusual phenomena, such as the broadening of the Compton emission spectrum \cite{Khalaf}, high-order above-threshold ionization spectrum \cite{Wangshi} and HHG spectrum \cite{Gorlach}, the modification of electronic trajectories in HHG \cite{Tzur,Dean} and strong-field ionization \cite{Fang}, and the generation of displaced squeezed state in HHG \cite{Theidel}. 
Very recently, HHG in solids driven by BSV has been demonstrated in experiment \cite{Rasputnyi,Lemieux}. 
Nowadays, the study of HHG under quantum light remains in its infancy stage.
The influence of photon statistics of the driving field on HHG features still merits further exploration.

\begin{figure*}[htb]
	\centerline{
		\includegraphics[width=17.5cm]{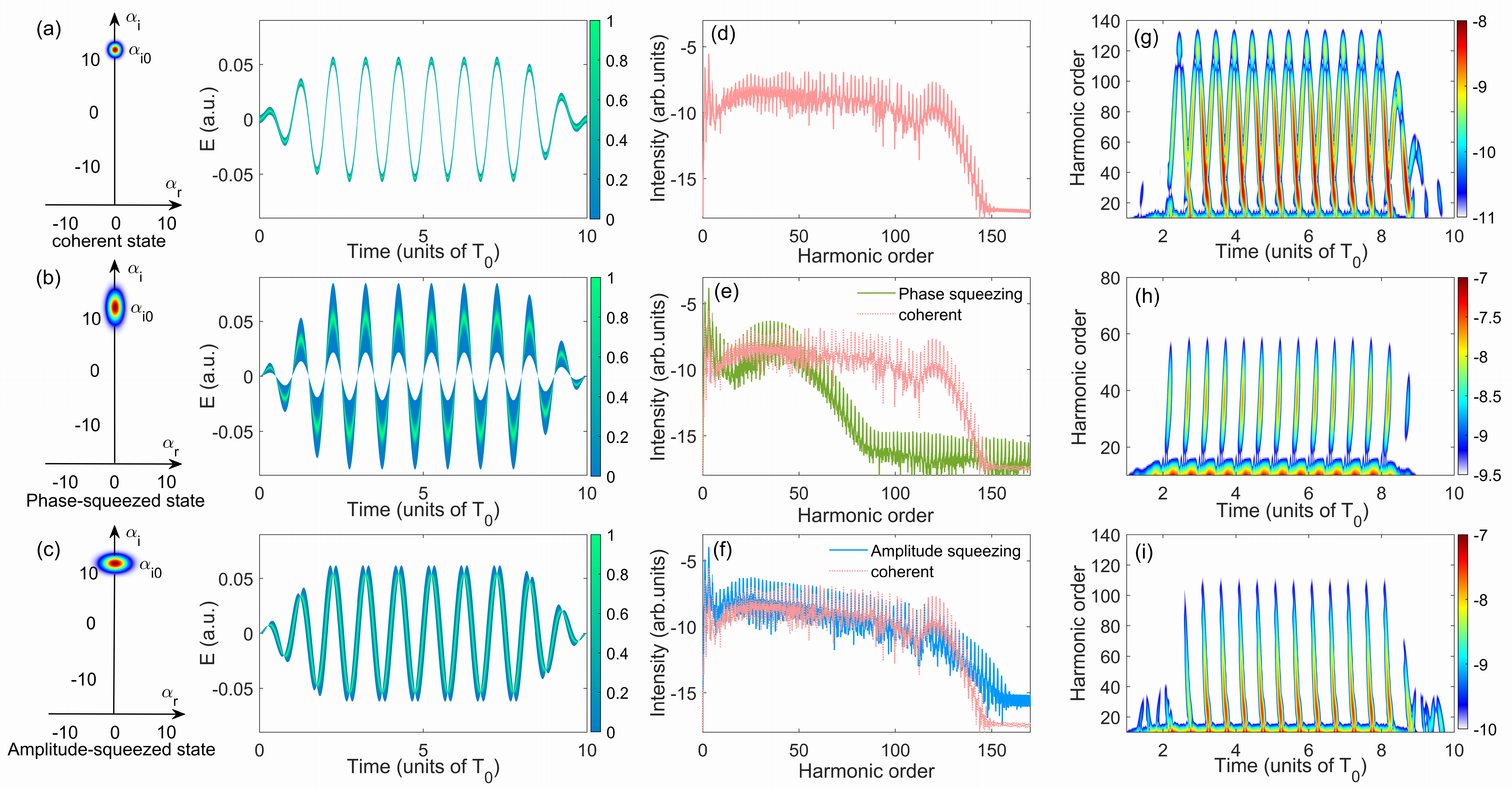}}
	\caption{(a)-(c) The quasiprobability distributions (left panels) and laser fields (right panels) of the coherent, phase-squeezed ($r=1$) and amplitude-squeezed ($r=-1$) states. (d)-(f) and (g)-(i) Harmonic spectra and corresponding time-frequency analysis obtained by fields in (a)-(c). For comparison, the harmonic spectrum driven by the coherent field is also presented (dotted line) in Figs. 1(e)-1(f).} 
\end{figure*}

In this work, we theoretically study the HHG process driven by squeezed lights and demonstrate that the quantum paths of  HHG  can effectively controlled by shaping the quantum state of the squeezed light. In traditional HHG studies by classical lights, the quantum path control is usually achieved by the phase-matching technique  \cite{80as,PG1,PR1,PR2,PR3,PR4}.
However, for different harmonics, the phase-matching conditions are usually different.
For given laser conditions, only a subset of harmonics within the plateau can be phase-matched for generating single-path harmonics, which will restrict the achievable duration of IAP.
Recently, the generation of ultrashort IAPs is still a fundamental challenge in attosecond science.
The shortest IAP experimentally achieved to date remains at 43 as \cite{Worner1}.
Driven by the phase-squeezed or amplitude-squeezed field, we find that the selection of specific quantum paths can be achieved in the whole harmonic plateau. 
This capability holds great potential for IAP generation with much shorter duration.

\begin{figure}[htb]
	\centerline{
		\includegraphics[width=9cm]{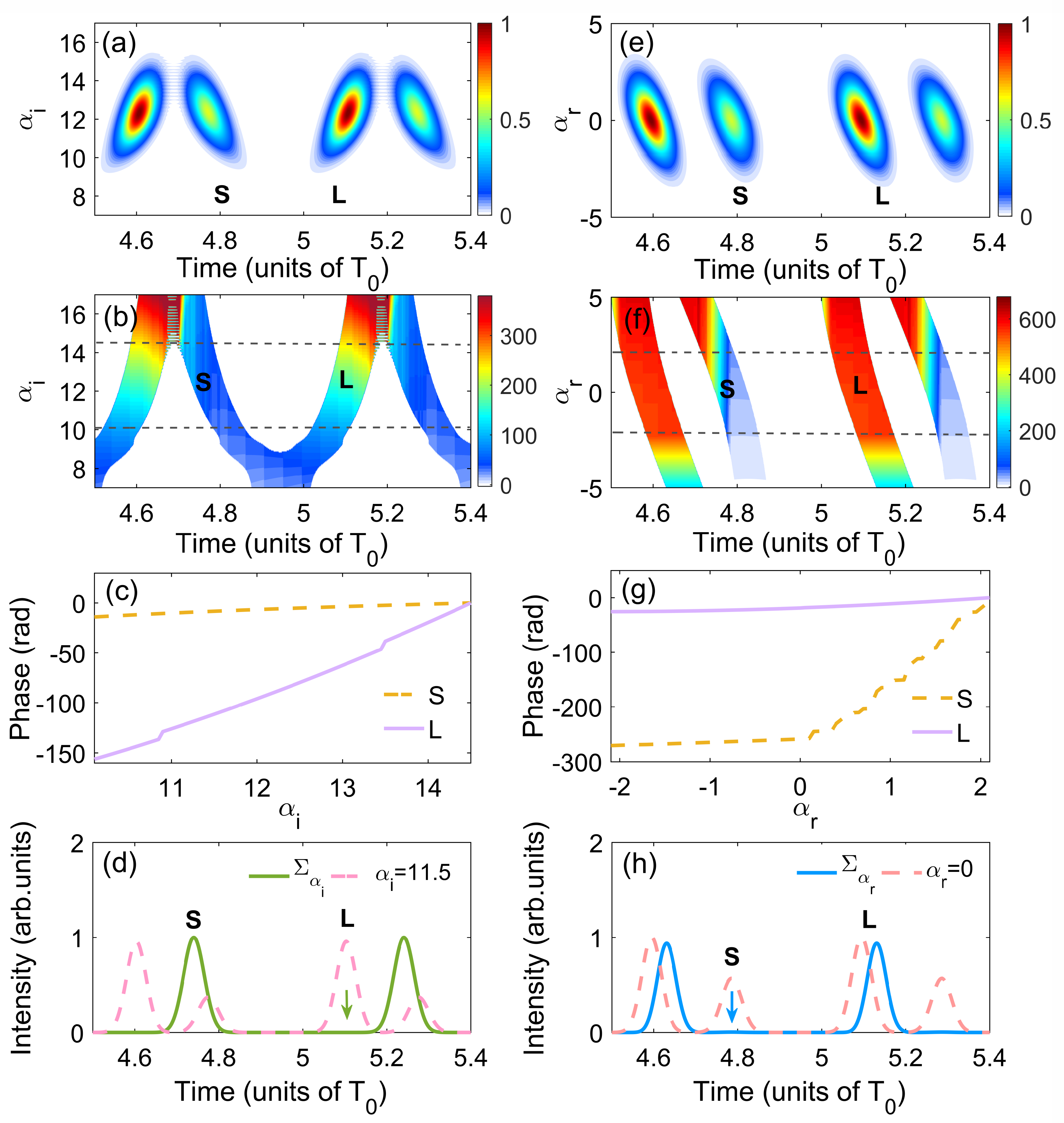}}
	\caption{(a)-(b) Time-dependent intensities and phases of H71 as a function of $\alpha_i$ in the phase-squeezed field with $r=1$. (c) Harmonic phase of H71 from the short and long paths as a function of $\alpha_i$ in the range from 10.1 to 14.5 [as marked by the dashed lines in (b)]. (d) Normalized time-dependent signals of H71 for $\alpha_i=11.5$ (dashed line) and that after coherent average over $\alpha_i$ (solid line). (e)-(h) Same as (a)-(d), but in the amplitude-squeezed field with $r=-1$.} 
\end{figure}

The interaction of quantum light with an atom can be described by the time-dependent Schr\"{o}dinger equation (TDSE) \cite{Gorlach,Tzur},
\begin{eqnarray}   
i\hbar\frac{\partial \rho(t)}{\partial t}=[(\hat{H}_0+\textbf{r}\cdot \textbf{E}+\hat{H}_f),\rho(t)],
\end{eqnarray}
where $\hat{H}_0=-\frac{1}{2m}\bigtriangledown^2+U(\textbf{r})$ is the Hamiltonian of the atom with $U(\textbf{r})$ the Coulomb potential.
$\hat{H}_f$ is the quantized electromagnetic field Hamiltonian.
$\bf{r}\cdot \bf{E}$ denotes the interaction between the laser and the atom.
$\rho(t)$ is the density matrix of the 
system.
At $t$=0, $\rho(0)=|0\rangle\langle0|\otimes \rho_f$.
$|0\rangle$ is the ground state of atom.
$\rho_f$ is the density matrix of the quantum light, which is given by \cite{Kim,Drummond}
\begin{eqnarray}
\rho_f=\int P(\alpha,\beta^*) \frac{|\alpha\rangle \langle\beta|}{\langle\beta|\alpha\rangle}d^2\alpha d^2 \beta,
\end{eqnarray}
where $|\alpha\rangle=|\alpha_r+i\alpha_i\rangle$ represents a coherent state of the quantum light. $\alpha_r$ and $\alpha_i$ are the real and imaginary parts of the coherent parameter $\alpha$.
The same for the notation $|\beta\rangle$.
$P(\alpha,\beta^*) = \frac{1}{4\pi}\mathrm{exp}[-\frac{\left| \alpha-\beta \right|^{2} }{4}]Q(\frac{\alpha+\beta}{2})$ is the generalized Glauber representation with $Q(\alpha)$ the Husimi quasiprobability distribution of the quantum light.
By inserting Eq. (2) into Eq. (1), the density matrix of electron can be obtained by
\begin{eqnarray}
\rho_{e} (t) = \int  P(\alpha,\beta^{\ast})\left| \phi_{\alpha}(t)\right\rangle \left\langle \phi_{\beta}(t)\right|d^{2}\alpha d^{2}\beta.
\end{eqnarray}
Here $|\phi_{\alpha,\beta}(t)\rangle$ is the time-dependent wavefunction of the electron, which can be obtained by solving the TDSE under the strong-field approximation (SFA) framework (see Appendix A).
Here, it is worth mentioning that the most accurate approach
for calculating HHG is by solving the TDSE.
To verify the accuracy of the SFA model, we have calculated the HHG spectra for a classsical driving field (as defined below) with both the TDSE and SFA methods in Appendix B.
The result shows that the harmonic spectrum obtained from the SFA model is in good overall agreement with that from the
TDSE model, indicating the validity of the SFA model.
   
The time-dependent dipole moment for HHG can be obtained by $z(t)=\rm{Tr}[z\rho_e(t)]=\int \textit{P}(\alpha,\beta^*)\langle\phi_\beta(t)|z|\phi_\alpha(t)\rangle d^2 \alpha d^2 \beta$. Here, the laser is assumed to be 
polarized along the $z$ direction.
Given that $P(\alpha,\beta^*)=P(\beta,\alpha^*)$ and $\int P(\beta,\alpha^{\ast}) d^{2}\beta= \int P(\alpha,\beta^{\ast}) d^{2}\beta$, the time-dependent dipole moment then can be rewritten as,
\begin{eqnarray}
z(t)=\int P(\alpha) z_\alpha(t) d^2 \alpha,
\end{eqnarray}
where $z_\alpha(t)=\langle\phi_{\alpha}(t)|z|\phi_{\alpha}(t)\rangle$
is the expectation value of the dipole moment for the coherent state $|\alpha\rangle$.
$P(\alpha)=\int P(\alpha,\beta^*)d^2\beta$ is the quasiprobability distribution of the quantum field. For a given squeezed coherent state $|\alpha_0,r\rangle$, $P(\alpha)$ can be found in \cite{Tzur}.
Here $r$ is the squeezing parameter.
$r=$ 0, $r>$ 0 and  $r<$ 0 correspond to the coherent, phase-squeezed and amplitude-squeezed states, respectively.

The harmonic spectrum in the quantum light is then calculated by the Fourier transform of the time-dependent dipole acceleration, i.e., $a_q=\int a(t) e^{-iq\omega t} dt$, where $a(t)$=\"{z}$(t)$ is the time-dependent dipole acceleration, $q$ is the harmonic order, and $\omega$ is the laser frequency. By superposing the harmonics within a given spectral range, the attosecond pulse can be obtained by $I_{atto}=|\sum_q a_q e^{iq\omega t}|^2$. 

For a coherent state $|\alpha\rangle$, the classical component of the laser field can be expressed as  $E_{\alpha}(t)=f(t)[2\hbar\omega/(\varepsilon_0V)]^{1/2}[\alpha_rcos(\omega t)+\alpha_isin(\omega t)]$.
$\varepsilon_0$ is the vacuum permittivity, $V$ is the quantization volume, and
$f(t)$ is the laser envelope.
In our simulations, we used a trapezoidal envelope with two-cycle rising and falling edges, and a six-cycle plateau.
The laser intensity and wavelength are $1\times10^{14}$ W$/$cm$^2$ and 1600 nm, respectively, and $V$ is set to $5\times10^{-27} m^3$. The left panels of Figs. 1(a)-1(c) display the quasiprobability distributions $P(\alpha)$ of the coherent, phase-squeezed and amplitude-squeezed states, corresponding to $r$= 0, 1 and -1, respectively. 
The phase-squeezed state is obtained by squeezing the coherent state along the horizontal direction, leading to the increased quantum fluctuation at the peaks and valleys of electric field [see right panel of Fig. 1(b)]. While the amplitude-squeezed state is obtained by squeezing the coherent state along the vertical direction. The quantum fluctuation is largest when the electric field approaches zero [see right panel of Fig. 1(c)].

In Figs. 1(d)-1(f), we simulate the harmonic spectra generated by the above electric fields with hydrogen atom. 
Notably, the structures of the harmonic spectra produced by these three fields differ significantly. Specifically, the harmonic cutoff is reduced in the phase-squeezed field compared to the coherent field.
For deeper insight, we have performed  time-frequency analysis of the harmonic spectra in Figs. 1(d)-1(f) using the Gabor transform,
\begin{eqnarray}
G(q,t_0)=\dfrac{1}{\sqrt{2\pi}}\int z(t) e^{-iq\omega t}e^{-\dfrac{(t-t_0)^2}{2\sigma^2}}dt,
\end{eqnarray}
where $\sigma$ is the width of the Gaussian window, which is set to $1/(5\omega)$ to balance the temporal and frequency resolutions.  
As shown in Fig. 1(g), in the coherent field, there are two branches within each half optical cycle, which are associated with the short and long quantum paths, contributing to harmonic emission.
This is consistent with the classical result \cite{Lewenstein}. 
While driven by the squeezed fields [Figs. 1(h)-1(i)], the coexistence of short and long paths disappears.
Only the short (long) quantum path survives in the phase-squeezed (amplitude-squeezed) field.



To understand the underlying physical mechanism of quantum path selection in HHG driven by squeezed fields, we turn to the SFA theory.
According to the SFA theory, the harmonic dipole moment of each quantum path driven by the coherent state $|\alpha\rangle$ can be expressed as $z_\alpha^{n}(q)= A_{\alpha}^{n}e^{i\varphi_{\alpha}^{n}}$.
Here the superscript $n=s,l$ denotes the short or long quantum paths, respectively.
$A$ and $\varphi$ represent the amplitude and phase of HHG from each quantum  path.                                                                                                                                                                                                                                                                                                                                                                                                                                                                                                                                                                                                                                                                                                                                                                                                                                                                                                                                   
In the squeezed field, the harmonic dipole moment of each path is a superposition of contributions from different coherent states, i.e, 
\begin{eqnarray}
z^{n}(q)=\int P(\alpha)z_\alpha^{n}(q) d^2 \alpha=\int P(\alpha) A_{\alpha}^{n}e^{i\varphi_{\alpha}^{n}}d^2 \alpha.
\end{eqnarray}
From Eq. (6), it is evident that the quasiprobability distribution $P(\alpha)$ of the squeezed field will lead to a distribution of both the amplitude and phase of HHG from each quantum path.

In Fig. 2, we take the 71st harmonic (H71) as an example to examine the harmonic amplitude and phase in squeezed fields.   
Figures 2(a) and 2(b) show the time-dependent intensities and phases of H71 as a function of $\alpha_i$ in the phase-squeezed field.
Here, $\alpha_r$ is fixed at 0, where the quantum fluctuation of phase-squeezed field is maximal.
As shown in Fig. 2(a),  for each coherent state $\alpha_i$, there are two burst events corresponding to the short and long quantum paths (labeled S and L) within each half optical cycle of the driving field, say 4.7-5.2 T$_0$ (where T$_0$ is the optical cycle of the laser).
This behavior is consistent with the result in Fig. 1(g).
As $\alpha_{i}$ changes, the short and long paths exhibit similar intensity distributions, except for the different maxima.
This indicates that the eventual intensities of the short and long paths are determined by their respective phase distributions.   
As shown in Fig. 2(b), the phases of the short and long paths have different behaviours as $\alpha_i$ changes.
For clarity, in Fig. 2(c), we have plotted the phases of the short and long paths within 4.7-5.2 T$_0$ with $\alpha_i$ varying from 10.1 to 14.5 where the harmonic intensity is dominant [see Fig. 2(a)].
Note that, for a multicycle trapezoidal driving laser, the harmonic emission within each half optical cycle is identical.
Here we mainly focus on the results within 4.7-5.2 T$_0$.
From Fig. 2(c), it's clear that the phase of the long path varies over a broader range than that of the short path as $\alpha_i$ changes.
According to the SFA theory, the phase of each quantum path  can be related to the action $S_\alpha(\textbf{p},t_i,t_r)$ accumulated by the electron in the laser field.
Here $\textbf{p}$ is the momentum of the electron, $t_i$ and $t_r$ are the ionization and recombination times of the electron.
The different electron motions in the laser field, i.e., the different ionization and recombination times, lead to the different phase distributions of the short and long paths in Fig. 2(c).
The larger phase uncertainty associated with the long path gives rise to the destructive interference, leading to the suppression of the long path.
To verify this, we have plotted in Fig. 2(d) the time-dependent signals of H71 for a single coherent state $\alpha_i=11.5$ (dashed line) and also that (solid line) after coherent superposition over all the states $\alpha_i$.
It's evident that for a single coherent state $\alpha_i$, the harmonic bursts from the short and long paths coexist within each half optical cycle. 
While after coherent superposition over $\alpha_i$, the emission from long quantum path is dramatically suppressed.

The selection of the long quantum path in the amplitude-squeezed field can be understood in a similar manner. 
Figures 2(e) and 2(f) show the time-dependent intensities and phases of H71 for different coherent states in the amplitude-squeezed field.
Here, $\alpha_i$ is fixed at 11.5.
Similar to the phase-squeezed field, HHG from the short and long paths in the amplitude-squeezed field also exhibit similar intensity distributions, but differ in their phase distributions as $\alpha_r$ changes. 
In the amplitude-squeezed field, the short path has a broader phase distribution as $\alpha_r$ changes as shown in Fig. 2(g), which plotted the phases of the short and long paths within 4.7-5.2 T$_0$ with $\alpha_r$ varying from -2.1 to 2.1. 
This broadening of the phase distribution for the short path results in destructive interference of the short-path harmonics from the different coherent states.
As a consequence, the short path is suppressed after coherent superposition over $\alpha_r$ [see solid line in Fig. 2(h)].

\begin{figure}[t]
	\centerline{
		\includegraphics[width=9cm]{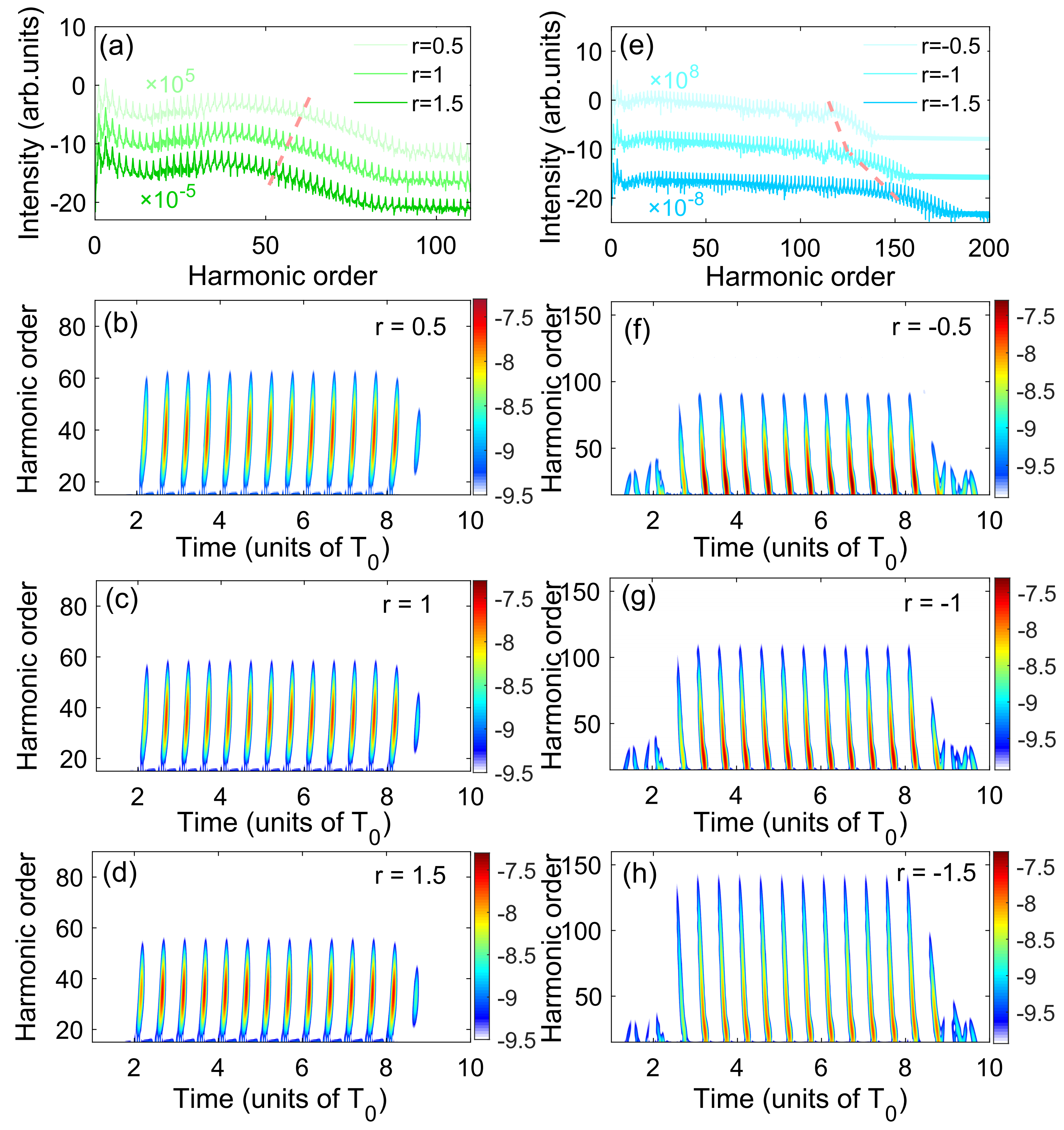}}
	\caption{(a)-(d) Harmonic spectra and corresponding time-frequency analysis driven by the phase-squeezed fields with r of 0.5, 1 and 1.5. (e)-(h) Same as (a)-(d), but for the amplitude-squeezed fields with r of $-$0.5, $-$1 and $-$1.5.} 
\end{figure}

We have also studied the influence of the squeezing parameter $r$ on the quantum path control. 
Figure 3(a) shows the harmonic spectra calculated for phase-squeezed fields with  $r$= 0.5, 1 and 1.5, respectively.
Figures 3(b)-3(d) present the corresponding time-frequency analysis of the harmonic spectra in Fig. 3(a).
As shown, the selection of the short quantum path in the phase-squeezed field remains effective for different squeezing parameters $r$. But the harmonic cutoff decreases as the squeezing parameter $r$ increases.
This is because higher harmonic orders experience poorer phase matching across the different quantum states of squeezed field as $r$ increases, leading to a more rapid decrease in harmonic intensity.
This is also the reason why the harmonic cutoff in the phase-squeezed field is lower than that in the coherent field, as seen in Fig. 1(e). 
For the amplitude-squeezed fields, the selection of the long quantum path is also maintained as the squeezing parameters $r$ varies [see Figs. 3(f)-(h)].
But the harmonic cutoff slightly increases as the squeezing parameter $r$ increases.
This is due to the slight increase in the maximum intensity of the amplitude-squeezed field as 
$r$ rises.
It is important to note that in experiment,  squeezed light can be generated by combining a coherent light with a BSV light \cite{Scully,Paris}.
In Appendix C, we have calculated the mean values and variances of the quadratures for the squeezed lights used above.
For a coherent light with an intensity of $1\times10^{14}$ W$/$cm$^2$ used here, the most advanced BSV light source currently available, with a peak intensity of approximately $1.5$$\times$$10^{12}$ W$/$cm$^2$ \cite{Finger}, can support the generation of a squeezed field with $|r|$$\approx$1. However, as demonstrated above, quantum path selection can still be achieved even with a squeezed field where $|r|$$<$1.

\begin{figure}[t]
	\centerline{
		\includegraphics[width=8.3cm]{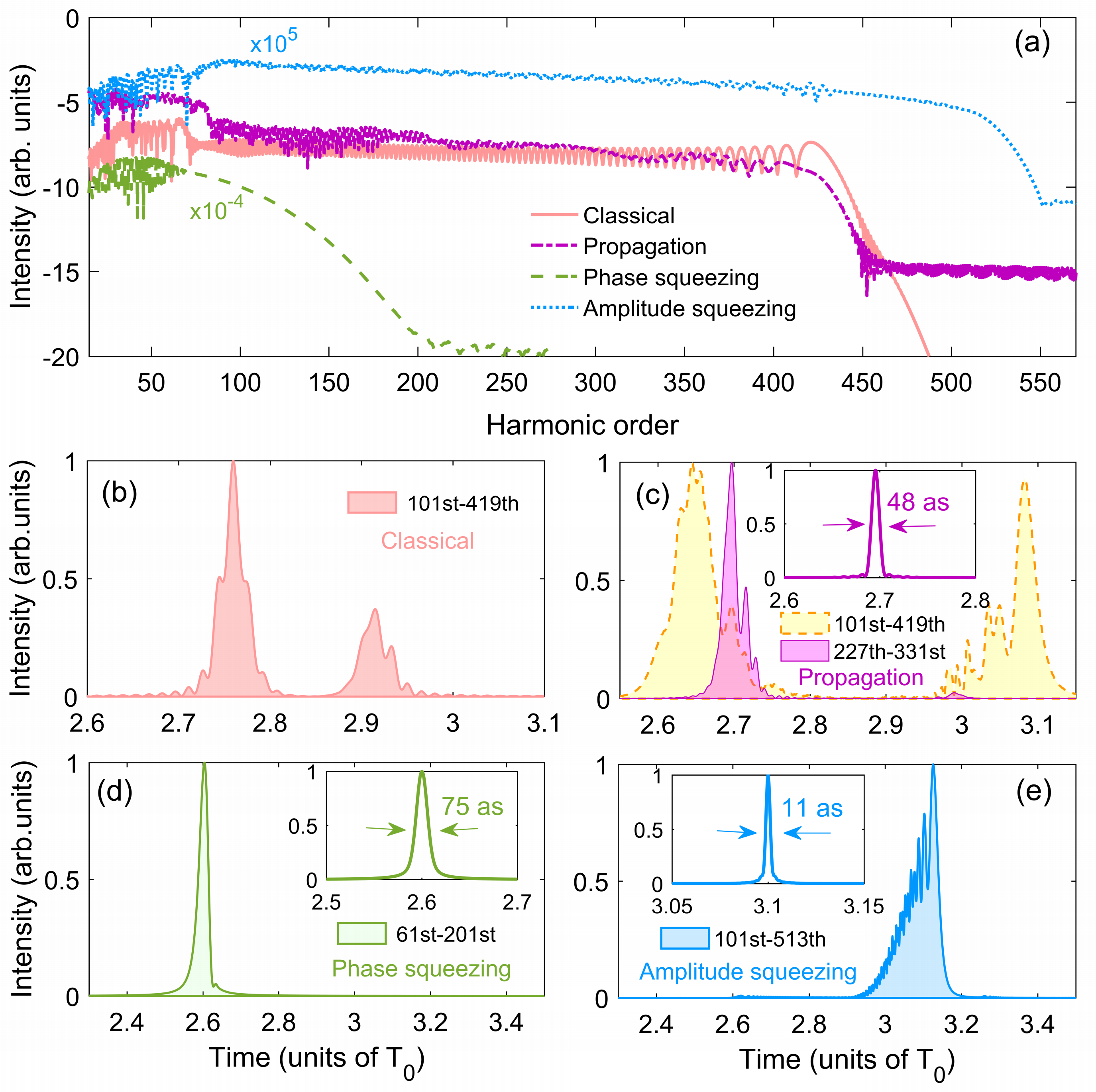}}
	\caption{(a) Harmonic spectra in a single-cycle phase-squeezed ($r$=0.01), and amplitude-squeezed ($r$=$-$1.5) fields. For comparison, the single-atom and macroscopic harmonic spectra in the classical field are also presented. (b)-(e) Normalized temporal profiles of attosecond pulse synthesized from the harmonic spectra in (a). The spectral range for IAP synthesis is given in the legends. The insets in (c)-(e) are the corresponding FTL pulses of the IAPs.} 
\end{figure}

A key application of the quantum path control in HHG is the generation of IAPs.
In squeezed fields, quantum path selection can be inherently achieved at the single-atom level. More importantly, this selection is effective across the entire harmonic plateau (see Figs. 1 and 3), extending the available spectral range for IAP synthesis and facilitating the generation of shorter IAPs.
To demonstrate this point, we employ a single cycle 1600-nm driving field to gate harmonic emission within a single half-optical cycle, enabling IAP generation. A higher laser intensity of 5$\times$$10^{14}$ W$/$cm$^2$ is used to broaden the harmonic plateau, while helium atom, with its high ionization potential, is chosen to suppress the ionization. Figure 4(a) presents the harmonic spectra driven by phase- and amplitude-squeezed fields. For comparison, the harmonic spectrum generated by the classical field is also shown. As expected, the classical field results in deep modulations in the supercontinuum region (101st to 419th harmonics) due to the interference between short and long quantum paths. Consequently, superposing these harmonics, a pair of attosecond pulse bursts with comparable intensity is produced [Fig. 4(b)]. In this case, additional macroscopic propagation is required to achieve single-quantum-path selection and then generate an IAP.
In Fig. 4(a), we have also simulated the propagation of the classical HHG by solving the three-dimensional Maxwell wave
equation \cite{He3,Feng,Jin}. 
The simulation considers a focused laser beam with a 30-$\mu$m waist and a 1-mm-long gas jet at a density of 100 Torr, positioned 1 mm after the laser focus---a typical experimental condition for short-path selection.
As shown in Fig. 4(a), the modulation depth is significantly reduced in the 227th-331st harmonic range, indicating that quantum path selection is effective within this region.
By superposing the entire supercontinuum (101st-419th), two attosecond pulses are still produced [dashed line in Fig. 4(c)]. However, when harmonics from only the 227th to 331st orders are used, an IAP is successfully generated [solid line in Fig. 4(c)]. Nonetheless, the restricted bandwidth sets a fundamental limit on the achievable Fourier-transform-limited (FTL) duration of the IAP, which is 48 as [see inset in Fig. 4(c)]. 

In contrast, when driven by a phase-squeezed field ($r$=0.01), a smooth harmonic supercontinuum spanning the 61st to 201st harmonics is generated [Fig. 4(a)]. This results in the direct production of an IAP associated with short-path emission around 2.6 T$_0$ [Fig. 4(d)]. However, due to the suppression of the cutoff region, this supercontinuum can only support an IAP with a FTL duration of 75 as [see inset in Fig. 4(d)]. For the amplitude-squeezed field ($r$=$-$1.5), a significantly broader harmonic supercontinuum, extending from the 101st to 513th harmonics, is obtained. With appropriate chirp compensation \cite{Worner1,53as}, an even shorter IAP, with an FTL duration of just 11 as, is generated at a later time, corresponding to long-path emission [see inset in Fig. 4(e)].

In conclusion, we have explored the quantum path control of HHG in the quantum-optical regime.
The simulation results indicate that the quantum paths of HHG  
can be effectively controlled by using different types of squeezed field. 
Specifically, the short (long) quantum path of HHG can be selected by using a phase (amplitude)-squeezed field. This feature is independent on the squeezing parameter. 
Based on the time-frequency analysis, the quantum path control of HHG is demonstrated to arise from the phase modulations induced by the quantum fluctuation of the squeezed fields.
More importantly, such a quantum path selection can be realized for harmonics in the whole plateau region.
Combining with traditional gating techniques \cite{Chang1}, our findings have the potential to break the current 43 as barrier and achieve the generation of IAP less than one atomic unit of time.
Such pulse permits studies of electron dynamics in gases, solids and liquids with unprecedented time resolution.
Looking forward, the inherent quantum path control of HHG in quantum light enables one-to-one mapping between harmonic photon energy and time, which will facilitate the development of high harmonic spectroscopy for probing the ultrafast electronic dynamics in quantum light.

\section*{Acknowledgement}
This work was supported by the National Key Research and Development Program of China (Grant No. 2023YFA1406800); National Natural Science Foundation of China (Grants No. 12104349, No. 62522505, No. 12474342, No. 92365106, No. 12225406, No. 12021004) and the Natural Science Foundation of Hubei Province (2022CFA039).

\section*{APPENDIX A: Quantum optical strong-field approximation (QSFA)}
In this section, we introduce the theoretical model used in our work for calculating the HHG in quantum light.
The interaction of a quantum light with an atom can be described by the TDSE \cite{Gorlach,Tzur},
\begin{eqnarray}   
i\hbar\frac{\partial \rho(t)}{\partial t}=[(\hat{H}_0+\textbf{r}\cdot \textbf{E}+\hat{H}_f),\rho(t)],
\end{eqnarray}
here $H_0=-\frac{1}{2m}\bigtriangledown^2+U(r)$ with $U(r)$ the atomic Coulomb potential.
$H_F=\hbar \omega$ \^{a}$^{\dagger}$\^{a} is the quantized electromagnetic field Hamiltonian.
$\omega$ is the frequency of the field.
\^{a}$^{\dagger}$ and \^{a} are the creation and annihilation operators, respectively.
$\textbf{E}=i\sqrt{\frac{\hbar \omega}{2\varepsilon_0 V}}$(\^{a}-\^{a}$^{\dagger}$) is the quantized electric field.
$\varepsilon_0$ is the vacuum permittivity.
$V$ is the quantum normalization volume.
$\bf{r}\cdot \bf{E}$ is the interaction term between the quantized electric field $\textbf{E}$ and atom in the dipole approximation.
$\rho(t)$ is the density matrix of the quantum light and electron. When $t=0$, $\rho(0)$ is equal to $|0\rangle \langle 0| \otimes \rho_f$.
$|0\rangle$ is the ground state of atom.
$\rho_f$ is the density matrix of the quantum light, which can be given by,
\begin{eqnarray}
\rho_f=\int P(\alpha,\beta^*) \frac{|\alpha\rangle \langle\beta|}{\langle\beta|\alpha\rangle}d^2\alpha d^2 \beta,
\end{eqnarray}
where $|\alpha\rangle=|\alpha_r+i\alpha_i\rangle$ and $|\beta\rangle=|\beta_r+i\beta_i\rangle$ are coherent states.
$P(\alpha,\beta^*) = \frac{1}{4\pi}\mathrm{exp}[-\frac{\left| \alpha-\beta \right|^{2} }{4}]Q(\frac{\alpha+\beta}{2})$ is the generalized Glauber $P$ representation with $Q(\alpha)$ the Husimi quasiprobability distribution of quantum light.
By inserting the Eq. (2) into Eq. (1),  we can obtain the density matrix of electron,
\begin{eqnarray}
\rho_{e} (t) = \int  P(\alpha,\beta^{\ast})\left| \phi_{\alpha}(t)\right\rangle \left\langle \phi_{\beta}(t)\right|d^{2}\alpha d^{2}\beta,
\end{eqnarray}
To obtain the harmonic spectrum, we calculate the expectation value of the time-dependent dipole moment \cite{Scully},
\begin{eqnarray}
z(t)=\rm{Tr}[z\rho_e(t)]=\int P(\alpha,\beta^*)\langle\phi_\beta(t)|z|\phi_\alpha(t)\rangle d^2 \alpha d^2 \beta, \nonumber\\
\end{eqnarray}
$|\phi_{\alpha,\beta}(t)\rangle$ is the time-dependent wavefunction of electron driven by the coherent state $|\alpha\rangle$ or $|\beta\rangle$, which can be obtained by solving the semi-classical TDSE,
\begin{eqnarray}
i\hbar\frac{\partial|\phi_{\alpha,\beta}(t)\rangle}{\partial t}=[H_0+z\cdot E_{\alpha,\beta}(t)]|\phi_{\alpha,\beta}(t)\rangle,
\end{eqnarray}
where $E_{\alpha}(t)$ is the classical field component of the coherent state $|\alpha\rangle$, which can be expressed as
$E_{\alpha}(t)=f(t)[2\hbar\omega/(\varepsilon_0V)]^{1/2}[\alpha_rcos(\omega t)+\alpha_isin(\omega t)]$ with $f(t)$ of the laser envelope.
The field $E_{\alpha}(t)$ is assumed to be polarized along $z$ direction.
The notation of $E_{\beta}(t)$ is the same.
Solving the Eq. (5) under the SFA framework \cite{Lewenstein}, we can obtain $|\phi_{\alpha,\beta}(t)\rangle=e^{\frac{i}{\hbar}I_p t}(a_0(t)|0\rangle+\int b_{\alpha,\beta}(\boldsymbol{\nu},t) \cdot |\boldsymbol{\nu} \rangle d^{3}\boldsymbol{\nu}$.
Here $I_p$ is the ionization potential of atom,
$a_0(t)$ is the ground-state amplitude,
$|\boldsymbol{\nu} \rangle$ represents the continuum states,
$b_{\alpha,\beta}(\boldsymbol{\nu},t)$ is the amplitude of the corresponding continuum state.
Neglecting the bound-bound and continuum-continuum transitions, we have  
\begin{eqnarray}
\left\langle \phi_{\beta}(t) \right| z \left| \phi_{\alpha}(t) \right\rangle & =& \int b_\alpha( \mathit{ \boldsymbol{\nu}},t) \cdot \langle0|z|  \boldsymbol{\nu} \rangle d^{3}  \boldsymbol{\nu} \nonumber\\  
&& +\int b^{\ast}_{\beta}(  \boldsymbol{\nu},t) \cdot \left\langle   \boldsymbol{\nu} \right| z \left| 0 \right\rangle d^{3} \boldsymbol{\nu}.
\end{eqnarray} 
Inserting Eq. (12) into Eq. (10), we can obtain,
\begin{eqnarray}
z(t) = \int d^{2}\alpha(\int P(\alpha,\beta^{\ast}) d^{2}\beta) \int b_\alpha( \boldsymbol{\nu},t) \cdot \langle0|z| \boldsymbol{\nu} \rangle d^{3} \boldsymbol{\nu} \nonumber\\  
+\int d^{2}\beta(\int P(\alpha,\beta^{\ast}) d^{2}\alpha)\int b^{\ast}_{\beta}( \boldsymbol{\nu},t) \cdot \left\langle  \boldsymbol{\nu} \right| z \left| 0 \right\rangle d^{3} \boldsymbol{\nu}. \nonumber\\
\end{eqnarray}
Due to  $P(\alpha,\beta^*)=P(\beta,\alpha^*)$ and $\int P(\beta,\alpha^{\ast}) d^{2}\beta= \int P(\alpha,\beta^{\ast}) d^{2}\beta$, Eq. (7) becomes,
\begin{eqnarray}
z(t)=\int d^{2}\alpha(\int P(\alpha,\beta^{\ast}) d^{2}\beta) \int b_\alpha(  \boldsymbol{\nu},t) \cdot \langle0|z| \boldsymbol{\nu} \rangle d^{3} \boldsymbol{\nu} +c.c.\nonumber\\
\end{eqnarray}
According to the expectation value of the dipole moment in a coherent state $|\alpha\rangle$, i.e.,  $z_{\alpha}(t)=\langle\phi_{\alpha}(t)|z|\phi_{\alpha}(t)\rangle=\int b_\alpha( \boldsymbol{\nu},t) \cdot \langle0|z| \boldsymbol{\nu} \rangle d^{3} \boldsymbol{\nu}+c.c.$\cite{Lewenstein}, we can obtain,
\begin{eqnarray}
z(t)=\int P(\alpha) z_\alpha(t) d^2 \alpha,
\end{eqnarray}
here $P(\alpha)=\int P(\alpha,\beta^*)d^2\beta$ is the quasiprobability distribution of the electric field.
For a given squeezed coherent state $|\alpha_0,r\rangle$, $r$ is the squeezing parameter, the Husimi $Q(\alpha)$ distribution is given by \cite{Kim}
\begin{eqnarray}
Q(\alpha)=\frac{1}{\pi\cosh(r)}e^{(-2\frac{(\alpha_{i}-\alpha_{i0})^{2}}{1+e^{2r}}-2\frac{(\alpha_{r}-\alpha_{r0})^{2}}{1+e^{-2r}})}.
\end{eqnarray} 
According to the relationship between $Q(\alpha)$ and $P(\alpha,\beta^*)$, we can obtain 
\begin{eqnarray}
P(\alpha,\beta^*)=e^{[ -\frac{(\alpha_{r}-\beta_{r})^{2}}{4}-\frac{(\alpha_{r}+\beta_{r}-2\alpha_{r0})^{2}}{2+2e^{-2r}}-\frac{(\alpha_{i}-\beta_{i})^{2}}{4}-\frac{(\alpha_{i}+\beta_{i}-2\alpha_{i0})^{2}}{2+2e^{2r}}]}. \nonumber\\
\end{eqnarray}
Then
\begin{eqnarray}
P(\alpha)=\dfrac{1}{\pi\cosh(r)}\dfrac{e^{-\dfrac{2(\alpha_{i}-\alpha_{i0})^{2}}{3+e^{2r}}-\dfrac{2(\alpha_{r}-\alpha_{r0})^{2}}{3+e^{-2r}}}(1+e^{2r})}{\sqrt{3+10e^{2r}+3e^{4r}}}.\nonumber\\
\end{eqnarray} 
$\alpha_0=\alpha_{r0}+i\alpha_{i0}$ is the complex amplitude of the squeezed coherent state $|\alpha_0, r\rangle$, representing the most probable position in $P(\alpha)$.

\section*{APPENDIX B: The comparison of HHG calculated with the SFA and TDSE models }
	Here, we calculate the harmonic spectra from a classical field with the three-dimensional TDSE and the SFA models, as shown in Fig. 5(a).
	In the simulations, we use a trapezoidal envelope with two-cycle rising and falling edges, and a six-cycle plateau.
	The intensity and wavelength of the classical field are $1\times10^{14}$ W$/$cm$^2$ and 1600 nm, respectively.
	One can see that the HHG spectrum obtained from the SFA model (pink solid line) is in good overall agreement with that from the TDSE model (blue dashed line).
	Furthermore, we have performed the time-frequency analysis for these two spectra [Figs. 5(b) and 5(c)].
	For the SFA result [Fig. 5(b)], two branches within each half optical cycle are observed, corresponding to the short and long quantum paths.
	This structure agrees well with the TDSE result [Fig. 5(c)], and is also consistent with previous report \cite{he22}.
	Note that the TDSE simulations naturally include contributions from multi-return trajectories, while our SFA treatment focuses on the conventional short and long paths and neglects multi-return contributions.
	Since HHG from multi-return trajectories is difficult to phase match, such contributions are expected to vanish after macroscopic propagation.
	Therefore, when restricting the analysis to the conventional short and long paths, the HHG spectra and the corresponding time-frequency structures obtained with the SFA and TDSE models show very good agreement, particularly in the plateau region.
	
	\begin{figure}[t]
		\centerline{
			\includegraphics[width=8.3cm]{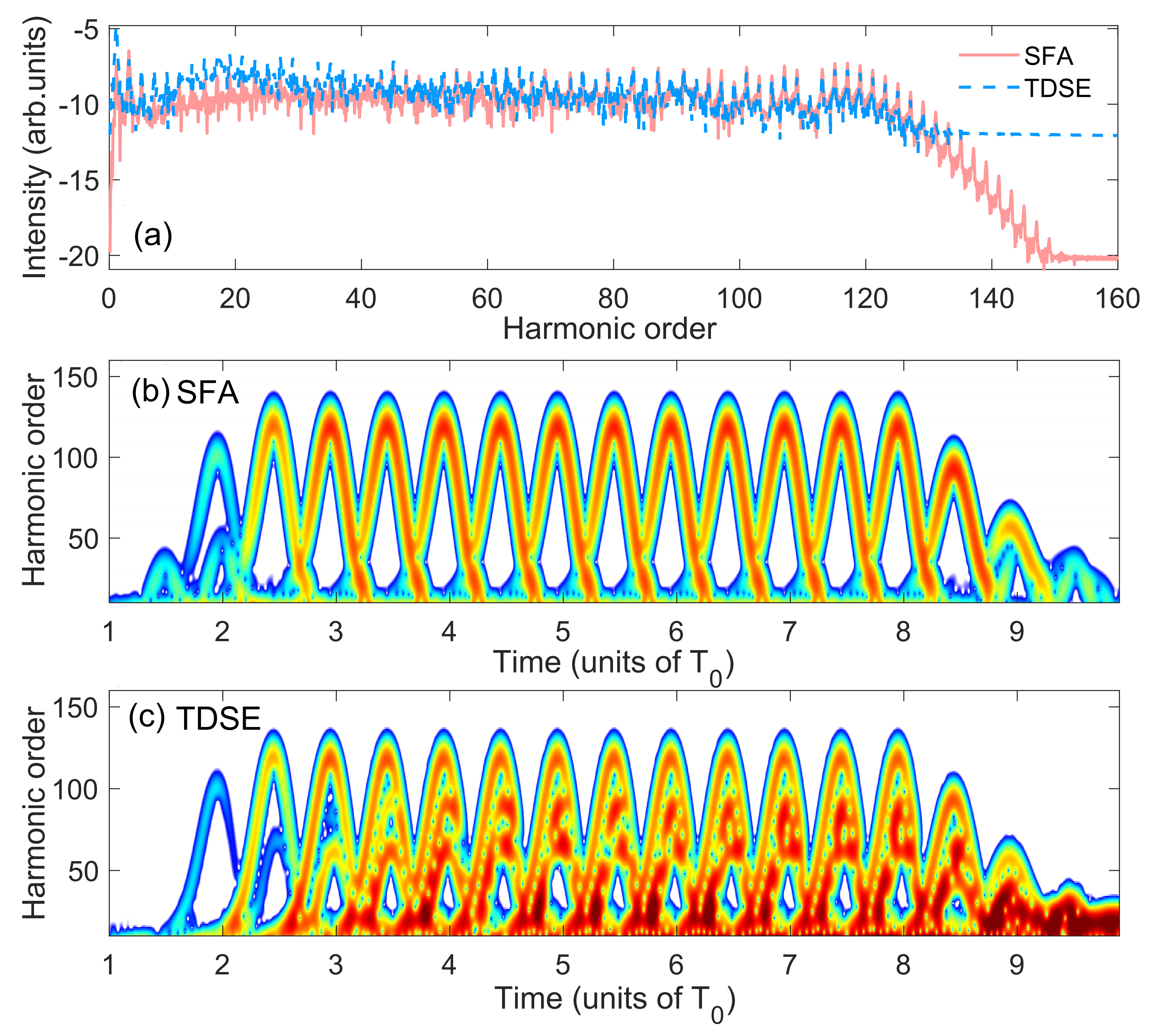}}
		\caption{(a) Harmonic spectra obtained using the SFA (pink solid line) and TDSE (blue dashed line) models. (b) Time-frequency analysis for the harmonic spectrum calculated with the SFA method. (c) Same as (b), but with the TDSE method. The intensity and wavelength of the classical field are $1\times10^{14}$ W$/$cm$^2$ and 1600 nm, respectively.} 
\end{figure}

\section*{APPENDIX C: The mean values and variances of the two quadratures of squeezed light}

Here we calculate the mean values and variances of the two quadratures of squeezed light.
For a given squeezed state $|\alpha_0,r\rangle$, the mean values of the two quadratures can be given by \cite{Wall}
\begin{eqnarray}
\langle Y_1 \rangle = \frac{1}{2}(\alpha_0 e^{-i\theta/2}+\alpha_0^* e^{i \theta /2}), \nonumber\\
\langle Y_2 \rangle = \frac{1}{2i}(\alpha_0 e^{-i\theta/2}-\alpha_0^* e^{i \theta /2}),
\end{eqnarray}
where $Y_1$ and $Y_2$ are the two quadratures along horizontal and vertical directions, respectively.  $\alpha_0=\alpha_{r0}+i\alpha_{i0}$, representing the most probable position in $P(\alpha)$.
$\theta$ is the squeezing angle.
The variances of the two quadratures of squeezed light are,
\begin{eqnarray}
V(Y_1)=\frac{1}{4}e^{-2\gamma},V(Y_2)=\frac{1}{4}e^{2\gamma},
\end{eqnarray}
where $r=\gamma e^{i\theta}$ is the squeezing parameter.
$\gamma$ is the squeezing amplitude.
In our work, $\alpha_{r0}=0$, $\alpha_{i0}=11.5$ a.u..
For the phase-squeezed field, $\theta$=0.
For the amplitude-squeezed field, $\theta$=$\pi$.
The corresponding mean values and variances of the two quadratures of squeezed light are shown in the TABLE 1.

In experiment,  squeezed light can be generated by combining a coherent light with a BSV light \cite{Scully,Paris}. 
For a coherent light with an intensity of $1\times10^{14}$ W$/$cm$^2$ used in the main text, the most advanced BSV light source currently available, with a peak intensity of approximately $1.5$$\times$$10^{12}$ W$/$cm$^2$ \cite{Finger}, can support the generation of a squeezed field with $|r|$$\approx$1.
To achieve squeezed light with $|r|$$>$1, further enhancement of the BSV light intensity is required.

\begin{table}[htbp]
	\centering
	\label{tab:my_table}        
	\begin{tabular}{|c|c|c|c|c|c|c|c|}  
		\hline	
		r & -1.5 & -1 & -0.5 & 0.01 & 0.5 & 1 & 1.5   \\
		\hline
		$ \left\langle Y_{1} \right\rangle $  & 11.5 & 11.5 & 11.5 & 0 & 0 & 0 & 0    \\
		\hline
		$\left\langle Y_{2} \right\rangle$  & 0 & 0 & 0 & 11.5 & 11.5 & 11.5 & 11.5  \\
		\hline
		$V(Y_{1})$ & 5.0214 & 1.8473 & 0.6796 & 0.2450 & 0.0920 & 0.0338 & 0.0124  \\
		\hline
		$V(Y_{2})$ & 0.0124 & 0.0338 & 0.0920 & 0.2551 & 0.6796 & 1.8473 & 5.0214  \\
		\hline
	\end{tabular}
	\caption{ The mean values and variances of the two quadratures of squeezed fields with squeezing parameters r =0.01, $\pm$0.5, $\pm$1, $\pm$1.5, respectively.} 
\end{table}

\end{document}